\documentclass[aps,apl,preprint,groupedaddress]{revtex4}
\usepackage{epsfig}
\usepackage{amsmath,amssymb}

\def\BibTeX{{\rm B\kern-.05em{\sc i\kern-.025em b}\kern-.08em
    T\kern-.1667em\lower.7ex\hbox{E}\kern-.125emX}}

\begin{document}

\title{Image acceleration in parallel magnetic resonance imaging by means of metamaterial magnetoinductive lenses}

\author{Manuel J. Freire$^1$}
\author{Marcos A. Lopez$^1$}
\author{Jose M. Algarin$^1$}
\author{Felix Breuer$^2$}
\author{Ricardo Marqu\'es$^1$}

\affiliation{$^1$Department of Electronics and Electromagnetism,
Faculty of Physics, University of Seville, Avda. Reina Mercedes
s/n, 41012 Sevilla, Spain } \email{freire@us.es}
\affiliation{$^2$Research Center Magnetic Resonance Bavaria (MRB),
W$\ddot{u}$rzburg, Germany}

\date{\today}

\begin{abstract}
Parallel magnetic resonance imaging (MRI) is a technique of image
acceleration which takes advantage of the localization of the
field of view (FOV) of coils in an array. In this letter we show
that metamaterial lenses based on capacitively-loaded rings can
provide higher localization of the FOV. Several lens designs are
systematically analyzed in order to find the structure providing
higher signal-to-noise-ratio. The magnetoinductive (MI) lens is
find to be the optimum structure and an experiment is developed to
show it. The ability of the fabricated MI lenses to accelerate the
image is quantified by means of the parameter known in the MRI
community as g--factor.

\end{abstract}


\maketitle

One of the most severe limitations of metamaterials for
applications is their narrow band response inherent to the
resonant nature of the elements that constitute the periodic
structure (see \cite{Marques-Book} and references therein). In
Magnetic Resonance Imaging (MRI), MR images are acquired by
measuring radiofrequency (RF) magnetic fields in the MHz range
inside a relatively narrow bandwidth of a few tens of kilohertz.
Therefore, the narrow band response of metamaterials is not a
problem for MRI, so that MRI should be then considered one of the
most promising field of applications for metamaterials. In
addition, since the wavelength associated with RF fields is of the
order of the meters, it is possible to use conventional printed
circuit techniques to develop quasi-continuous metamaterials with
constituent elements and periodicities two orders of magnitude
smaller than the wavelength. Several works have explored MRI
applications of metamaterials
\cite{Wiltshire}--\cite{Algarin-APL-2011-2} by using different
elements to build the periodic structure, such as swiss-rolls
\cite{Wiltshire}--\cite{Mathieu}, wires \cite{Radu} and
capacitively-loaded rings
\cite{Jelinek-JAP-2009}--\cite{Algarin-APL-2011-2}.
Capacitively-loaded rings have the key advantage over swiss rolls
and wires of providing three dimensional (3D) isotropy when they
are arranged in a cubic lattice, which is an essential property if
the device has to image 3D sources.

One of the most striking properties of metamaterials is the
ability of a metamaterial slab with relative permittivity
$\varepsilon$ and relative permeability $\mu$, both equal to $-1$,
to behave as a ''super-lens" with sub-wavelength resolution
\cite{Pendry-2000}. In the case of MRI applications, since the
frequency of operation is sufficiently low, we are in the realm of
the quasi-magnetostatics, and we only need a metamaterial slab
with $\mu=-1$ to behave as a ''super-lens" \cite{Pendry-2000}. In
previous works, some of the authors showed that a 3D array of
capacitively-loaded rings can behave as an effective homogeneous
medium with $\mu=-1$ \cite{Jelinek-JAP-2009}. The authors also
explored both theoretically and experimentally the ability of this
structure to behave as a ''super-lens" for MRI
\cite{Freire-APL-2008}--\cite{Algarin-NJP-2011}. Thus for example,
it was shown that in some circumstances, this device can enhance
the sensitivity of a single MRI surface coil, as a consequence of
its subwavelength focusing properties
\cite{Freire-APL-2008},\cite{Freire-JMR-2010}. As it is well
known, the generation of images in MRI is based on the detection
of spatial variations in the phase and frequency of the RF waves
absorbed and emitted by the nuclear spins of the imaged object.
These spatial variations are induced by some static magnetic field
gradients and the image acquisition involves many repeated
measurements and then signal processing (inverse Fourier
transforming) before obtaining an image of a single slice of
tissue. Actually, the long acquisition time is the main drawback
of MRI in comparison with computerized tomography, and time
reduction without degrading the signal-to-noise ratio (SNR) is the
main aim of research in the MRI comunity. Image acceleration in
MRI is achieved by means of techniques known in general as
parallel MRI (pMRI) \cite{PILS,SENSE,GRAPPA}. pMRI works by taking
advantage of the spatially sensitive information inherent in a
receiving array of multiple surface coils in order to partially
replace time-consuming spatial encoding. PILS, SENSE and GRAPPA
are examples of these parallel techniques
\cite{PILS,SENSE,GRAPPA}. For instance, in the PILS technique
\cite{PILS}, it is assumed that each individual coil in the array
has a localized sensitivity pattern or "field of view" (FOV).
However, this localization takes place only at distances very
close to the array because of the spreading of the magnetic field
at farther distances. SENSE and GRAPPA are the commercially
available techniques \cite{SENSE, GRAPPA}. These techniques are
not restricted to linear coil configurations or localized
sensitivities. However, coil sensitivity variation in the
phase-encoding direction in which the reduction is performed must
be ensured. In general, the ratio between the SNR in the
accelerated image after parallel imaging reconstruction
(SNR$_{acce}$) and the SNR of a full or conventional acquisition
(SNR$_{full}$) decreases with the square root of the reduction or
acceleration factor $R$ (for example, $R=2$ means that the
acquisition time reduces to one half) as well as an additional
coil geometry dependent factor known as geometry g--factor in the
parallel imaging community \cite{SENSE,Breuer}:
\begin{equation}
\frac{{\rm SNR}_{full}}{{\rm SNR}_{acce}}=g \cdot \sqrt{R}
\end{equation}
The g--factor results in a spatially variant noise enhancement
that strongly depends on the encoding capability of the receiver
array. It is worth to mention for the discussion, that overlapping
of the FOVs of adjacent coils in the array degrades the SNR of the
image in the overlapping region due to the noise correlation
between the coils \cite{GRAPPA}. This overlapping can be
quantitatively estimated by means of the g--factor. A smart design
of the array can minimize the g-factor in the overlapping region,
which means that the image can be accelerated (higher $R$) without
degrading much the SNR. Localizing the FOV results in a reduction
of the g--factor, and therefore, the image can be accelerated
without degrading the SNR.

In a previous work, the authors suggested that a metamaterial
''super-lens" can provide a time reduction in the acquisition
process if the imaging ability of this device is combined with the
encoding process of parallel techniques \cite{Freire-JAP-2008}.
This suggestion was numerically \cite{Freire-JMR-2010} and
experimentally \cite{Algarin-APL-2011} investigated by the
authors. The authors experimentally shown \cite{Algarin-APL-2011}
that a $\mu=-1$ slab consisting of a 3D array of
capacitively-loaded rings can help to discriminate the fields
produced by the coils at deeper distances inside the patient body,
so that this device could be advantageously used in pMRI
techniques in order to obtain a better localization of the FOV.
Although the reported device \cite{Algarin-APL-2011} actually
improved the localization of the FOV, the authors also realized
that the SNR was degraded in the full FOV by the presence of the
lens due to the additional ohmic losses of the device
\cite{Algarin-APL-2011}. Therefore, in order to achieve a
practical application in pMRI, further research aimed to a
reduction of these losses was required. In the present work,
capacitively-loaded ring lenses with different structures have
been investigated in order to look for a device providing higher
SNR. This research is carried out by means of a computational tool
developed by the authors for the calculation of the SNR provided
by MRI coils in the presence of capacitively-loaded ring
structures and a conducting phantom resembling human tissue
\cite{Algarin-NJP-2011}. This computational tool was previously
checked by the authors with experimental results
\cite{Algarin-NJP-2011}. Using this method, several structures
have been numerically and experimentally analyzed and an optimum
structure has been found. An MR experiment is shown to prove that
this optimum structure can provide image acceleration without
degrading the SNR, so that it would be suitable for a practical
application.

The configuration under analysis is shown in Fig.1. It consists of
a two-channel array of squared coils with a metamaterial structure
placed between these coils and the phantom. In a previous work
\cite{Algarin-APL-2011}, a similar configuration was analyzed. In
this configuration the metamaterial lens had a larger area than
the array. In the present research, we have found that the noise
coming from the metamaterial structure is reduced significantly if
the metamaterial lens is divided into two smaller lenses, each one
of them with an area smaller than the area of each coil (see Fig.
1). The magnetic field produced by a coil has a central main lobe
and side lobes (see Fig. 2), with side lobes corresponding to the
magnetic field vortex around the conducting loop. If the magnetic
field produced by the coil is decomposed into spatial Fourier
harmonics, the main lobe will be represented by low harmonics
whereas the side lobes will be represented by high harmonics,
corresponding to the strong spatial variations of the field. The
transfer function of split-ring metamaterial lenses
\cite{Jelinek-JAP-2009,Algarin-MET-2011} has a cutoff wavenumber
which prevents transferring of high harmonics, so that side lobes
are not transferred by the lens. Moreover, high harmonics related
with side lobes account for high losses in the lens, thus
increasing noise. Therefore, it is convenient to make the area of
the lens smaller than the area of the coil in order to reduce such
noise. Moreover, since the side lobes are the dominant source of
the noise correlation \cite{Breuer} between adjacent coils in an
array, it is also convenient for pMRI applications to transfer
only the main lobes. When the lenses are present they transfer the
main lobe but not the side lobes, so that the side lobes attenuate
in air and do not reach the phantom. In absence of the lenses, the
coils are closer to the phantom and the side lobes penetrate it.
The distance between the coils, the lenses and the phantom can be
optimized using the previously reported method
\cite{Algarin-NJP-2011} in order to get higher SNR.  In our
analysis, $\mu=-1$ lenses corresponding to 3D arrays of two and
one unit cells in depth were studied. A second type of lens
proposed in the past by the authors
\cite{Freire-JAP-2008,Freire-APL-2005,Freire-JAP-2006} and termed
magnetoinductive (MI) lens was also studied. The MI lens consists
of a pair of parallel 2D arrays of rings. The principle of
operation of the MI lens is different from the $\mu=-1$ lens. In
the MI lens, the operating frequency does no correspond to an
effective value of permeability but to the frequency between two
resonances \cite{Freire-JAP-2006} which are similar to plasmons in
negative permittivity devices \cite{Pendry-2000}. Moreover,
whereas the $\mu=-1$ lens is isotropic, the MI lens is anisotropic
since it only interacts with fields which are perpendicular to the
arrays. This is not a problem since the field produced by MRI
coils at closer distances is mainly axial. In our analysis, it was
found that the MI lens provides the higher SNR due to the lower
ohmic losses introduced by the structure.

Fig. 3 shows the computation of the resistance in a squared coil
of 12 cm in length in the presence of different lenses: a $\mu=-1$
lens with two unit cells in depth, a $\mu=-1$ lens with one unit
cells in depth and a MI lens. All these lenses are 90 mm in length
with $6 \times 6$ unit cells and periodicity 1.5 cm. Dimensions of
the rings are the same as in a previously reported device
\cite{Freire-APL-2008}, i.e., the rings are 4.935 mm in radius and
have 2.17 mm of strip width. The two arrays in the MI lens are
separated by a distance of 11 mm. The results in Fig. 3.a show
that the MI lens provides the lower resistance at the frequency of
63.6 MHz corresponding to the operating frequency of a 1.5T MRI
system. Fig. 3.b checks this results by comparing the theoretical
prediction for the MI lens with measurements carried out with an
Agilent PNA series E8363B Automatic Vector Network Analyzer.
Finally, Fig. 3.c shows the computation of the axial profile of
the SNR provided by all these lenses, the MI lens being the
structure providing the highest values at all distances.

Thus, two MI-lenses with $6 \times 6$ rings were designed and
fabricated to operate at 63.63 MHz for experiments in a 1.5 T
clinical MRI scanner. Each ring is 4.935 mm in radius and have
2.17 mm of strip width and contains a 470 pF non-magnetic
capacitor. The two arrays in the MI lens are separated by a
distance of 11 mm, and the distance between the coils and the
lenses and between the lenses and the phantom was 6 mm. Two
receive-only arrays with two 12x12 cm$^2$ elements were built. One
array has been combined with the MI-lenses. The elements in both
arrays are decoupled using a shared conductor with a decoupling
capacitor. Each element in both configurations was tuned to 63.63
MHz and matched to 50 $\Omega$ in presence of an agar--phantom
($\varepsilon$=90 and $\sigma$=1.2 S/m). The elements in the
arrays were also actively decoupled by a tuned trap circuit
including a PIN diode in transmission. The isolation achieved
between the elements in both setups was better than -30 dB. The
active decoupling by the traps has been found to be better than
-30 dB. In order to investigate the SNR performance of the arrays,
quantitative SNR maps were calculated from a series of identical
phantom images \cite{Ohliger} for both setups using a
gradient-echo sequence (parameters: TR= 500 ms, TE= 10 ms, FOV:
380 x 380 mm2, matrix: 256 x 256, slice thickness= 5 mm). All MR
images were acquired in the 1.5 T whole body system (Symphony
Magnetom, Siemens, Germany) sited at the University Hospital
Virgen Macarena (Seville, Spain).  Figure 4.a shows a comparison
of the calculated SNR-maps. In the presence of the MI lenses, the
distance between the coils and the phantom is 23 mm (the thickness
of the lens is 11 mm), the main lobes are transferred but not the
side lobes, so that the side lobes attenuate in air and do not
reach the phantom. In absence of the lenses, the coils are at 6 mm
from the phantom and the side lobes penetrate it. The results if
Fig. 4.a make apparent the ability of the MI lenses to localize
the FOV of the coils in the array. Moreover, for short distances,
the SNR provided by the lenses is even higher than in absence of
the lenses. This is a clear improvement in comparison with the
experimental results previously reported by the authors for a lens
larger than the array \cite{Algarin-APL-2011}. In addition, in
order to investigate the parallel imaging capabilities of the MI
lenses, GRAPPA reconstructions have been carried out
\cite{GRAPPA}. Corresponding GRAPPA g-factor maps \cite{Breuer}
and the noise correlations were calculated for a reduction factor
$R=3$. Figure 4.b shows a comparison of the GRAPPA g-factor maps.
The g-factor obtained in the space between the FOVs when the MI
lenses are present is clearly lower than in absence of the lenses,
so that with the MI lenses the acquisition can be accelerated
without degrading the SNR in this region. Therefore, MI lenses can
find a real application in MRI as devices which allows to
accelerate the image acquisition, thus providing a real benefit
for patients.

\newpage
\noindent {ACKNOWLEDGEMENTS}
\bigskip

This work has been supported by the Spanish Ministerio de Ciencia
e Innovacion under projects Consolider-EMET CSD2008-00066 and
TEC2010-16948 (SEACAM) and by the Spanish Junta de Andalucia under
project TIC-06238 (METAMED). We want also thank Dr. Carlos
Caparros, from University Hospital Virgen Macarena (Seville,
Spain), for his advice.

\newpage

\noindent {CAPTION TO FIGURES}

\bigskip

\noindent Figure 1. a): Sketch of a 3D $\mu=-1$ lens with two unit
cells in depth and a MI lens, both of them with an area of
$6\times 6$ unit cells. b): Photographs of the real devices
sketched in a). c): Sketch Configuration under analysis and
consisting of two lenses placed between a two-channel array of
coils and a phantom. d): Photograph of the real configuration with
MI lenses.

\bigskip

\noindent Figure 2. a): Magnetic field lines of a circular coil of
radius $R$. b): Plot of the axial field of the same coil at an
axial distance of $R$.

\bigskip

\noindent Figure 3. a): Calculation of the input resistance of a
coil of 12 cm in diameter at 15 mm from 3D $\mu=-1$ lenses of
two-unit cells in depth, one unit-cell in depth and a MI lens. b):
Comparison between calculation and measurement for the input
resistance of the MI lens.

\bigskip

 \noindent Figure 4. (a) SNR maps and (b) g--factor maps of an agar phantom for a
 two-channel array of coils with and without MI lenses.

\newpage

\begin{figure}[h]
\begin{center}
\includegraphics{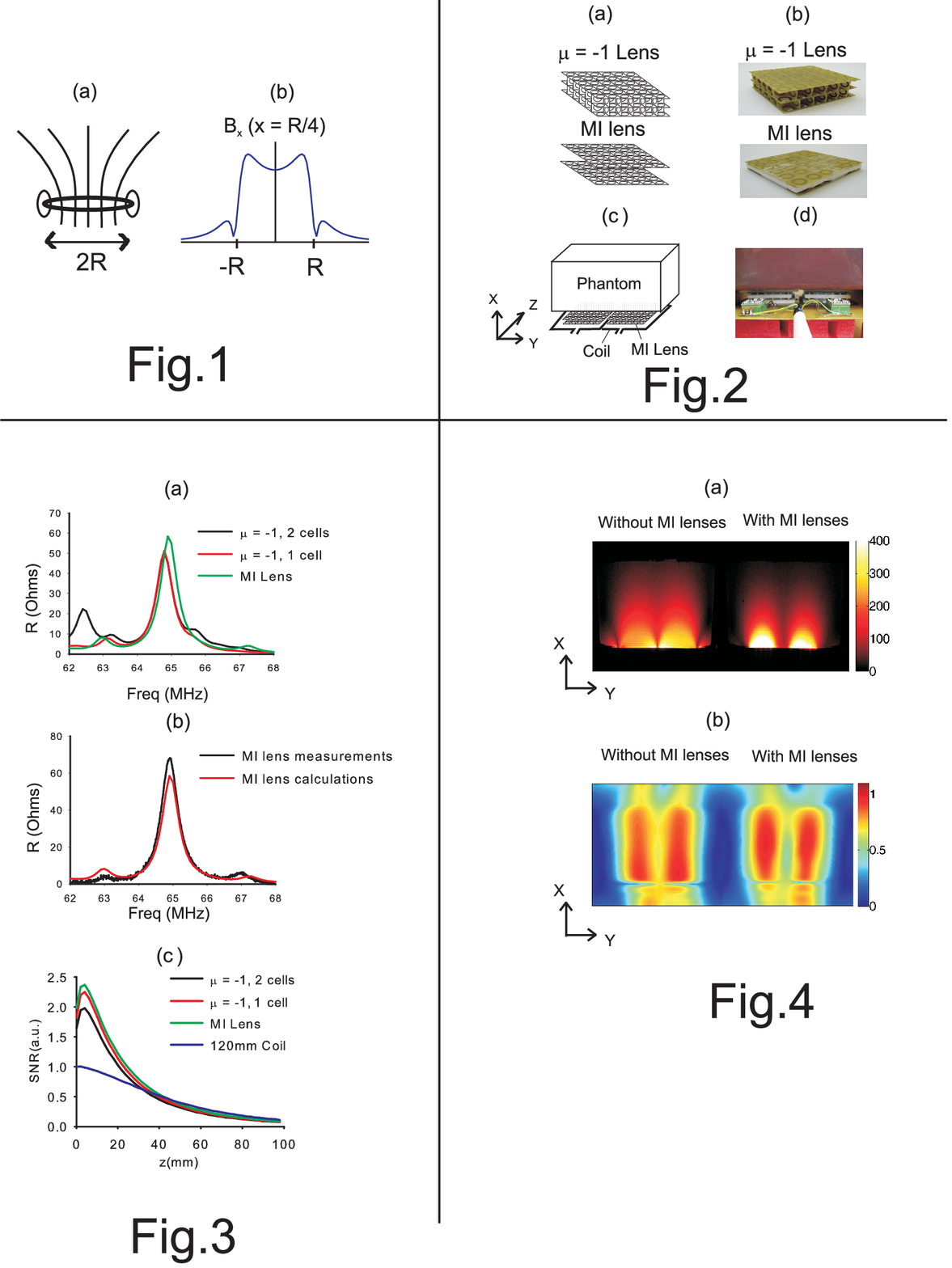}
\end{center}
\end{figure}

\end{document}